\documentclass[preprint,preprintnumbers,showpacs,amsmath,amssymb]{revtex4-1}
\usepackage{graphicx}
\usepackage{dcolumn}
\usepackage{bm}

\begin{document}

\title{Finite-temperature decoherence of spin states in a $\texttt{\{}\emph{Cu}_3\texttt{\}}$ single molecular magnet}

\author{Xiang Hao}

\author{Xiaoqun Wang}
\affiliation{Department of Physics, Renmin University of China, Beijing 100872, China}

\author{Chen Liu}

\author{Shiqun Zhu}
\affiliation{School of Physical Science and Technology, Soochow
University, Suzhou, Jiangsu 215006, People's Republic of China}

\begin{abstract}

We investigate the quantum evolution of spin states of a single molecular magnet in a local electric field. The decoherence of a $\texttt{\{}\emph{Cu}_3\texttt{\}}$ single molecular magnet weakly coupled to a thermal bosonic environment can be analyzed by the spin-boson model. Using the finite-temperature time-convolutionless quantum master equation, we obtain the analytical expression of the reduced density matrix of the system in the secular approximation. The suppressed and revived dynamical behavior of the spin states are presented by the oscillation of the chirality spin polarization on the time scale of the correlation time of the environment. The quantum decoherence can be effectively restrained with the help of the manipulation of local electric field and the environment spectral density function. Under the influence of the dissipation, the pointer states measured by the von Neumann entropy are calculated to manifest the entanglement property of the system-environment model.

\end{abstract}

\maketitle

\section{\label{sec1}Introduction}

The unavoidable interactions of all open quantum systems with environments often result in the dissipation and decoherence \cite{Breuer01, Weiss99}. Due to the exchange of energy and information between the system and the environment, the non-Markovian dynamics of quantum states always occur in the realistic experimental systems \cite{Hoeppe, Guo, Kuhr}. Recently, much attentions have been paid to the control of the decoherence of many-body quantum systems \cite{Carle, Ardavan} such as spin clusters and single molecular magnets \cite{Gatteschi}. As a class of systems with rich quantum properties, single molecular magnets at low energies can serve as a large-spin system or a collection of interacting spins \cite{Friedman, Thomas, Wernsdorfer}. These solid quantum spin systems are considered as promising carriers of quantum information \cite{Leuenberger, Lehmann}. Single molecular magnets with antiferromagnetic spin couplings can provide low-energy states for performing quantum logic gates \cite{Geogeot, Timco, Candini, Luis11}. Quantum decoherence is manifested when single molecular magnets are coupled to a spin bath \cite{Szallas, Morello, Prokofev, Coish}. The dissipation and decoherence always depend on the properties of the environment which can be described by a certain spectral density function \cite{Biercuk, Maniscalco, Haikka, Goan}. Therefore, a reasonable quantum manipulation method is necessary. At present, chemical manipulation can offer an efficient way to engineer intermolecular couplings and allow for interactions between qubits \cite{Troiani}. The decoherence from the chemical control cannot be easily eliminated because of the permanent interactions with the surrounding \cite{Szallas}. Simultaneously, the most straightforward and conventional way is to adopt an external magnetic field produced by electron spin resonance pulses \cite{Ardavan}. Although the decoherence of single molecular magnets can be suppressed by strong magnetic fields, it is preferable to apply electric fields that are controllable and suitable on very small spatial and temporal scales. The fast and precise quantum manipulations are helpful to the efficient realization of quantum logic gates. It is possible to apply time-dependent strong electric field close to single molecular magnet via a scanning tunnel microscopic tip \cite{Trif, Hirji06}. It has been found out that an electric field can be coupled to low-energy spin states of different chirality due to the absence of spin inversion symmetry in some single molecular magnets, such as $Cu_{3}$ \cite{Trif} , $V_{15}$ \cite{Chiorescu}, $Co_{3}$ \cite{Juan}, $Dy_{3}$ \cite{Luzon} and $Mn_{12}$ \cite{Friedman, Thomas}, etc. The effective spin electric coupling relies on the detailed structure of single molecular magnets at low energies \cite{Trif10}. Moreover, the use of microwave cavities can contribute to the indirect generation of fully controllable and long-range interaction between any two molecular magnets. This scheme based on electric-field local control can open up the possibility of scalable solid quantum information processing. Thus, the control of the decoherence of single molecular magnets driven by local electric field in low-temperature environments needs to be further studied.

This paper is organized as follows. In Sec. II, the time evolution of low-energy spin states in an electrically driven single molecular magnet without the spin inversion symmetry is studied by the spin-Boson model. As a physical example, a single molecular magnet $Cu_{3}$ is considered. The time-convolutionless non-Markovian master equation is obtained under the assumption of weak interactions with a thermal bosonic environment. Utilizing the secular approximation which means that the characteristic time scale of the system is much shorter than the correlation time of the environment, we derive the reduced density matrix of the low-energy spin states. In Sec. III, the decoherence of spin states can be shown by the dynamical behavior of the chirality spin polarization. We consider the reasonable Lorentzian environmental spectral density function characterizing the thermal bath. With respect to the entanglement between the system and environment, the pointer states measured by the von Neumann entropy are also investigated. Finally, a simple discussion concludes the paper.

\section{\label{sec2}Model of an open single molecular magnet in local electric field}

\subsection{\label{sec2:level21}Effective spin Hamiltonian}

Molecular magnets have clear features of coherent behavior and a variety of effective low-energy spin Hamiltonian is used for encoding qubits and implementing spin-based quantum computation. We can numerically investigate the low energy spectrum of a single molecular magnet by means of the Hubbard model~\cite{Zhao06}. A local electric field $\vec{\epsilon}(t)\thicksim 10^8 V/m$ can be applied to couple low-energy spin states of opposite chirality by means of the STM tips~ \cite{Trif10}. It is shown that both spin-orbit interactions and the absence of spin inversion symmetry can induce the the electric dipole matrix element $\vec{d}$ which is an important quantity in the effective spin-electric coupling. The strength of spin-electric coupling can be calculated by means of {\it ab initio} methods. For a single molecular magnet with an effective spin-electric coupling, the effective low-energy spin Hamiltonian is written as
\begin{equation}
\label{eq:1}
H^{eff}=H_{0}+H_{\epsilon}~,
\end{equation}
where $H_{0}$ is the low-energy spin Hamiltonian without the electric field and the effective spin-electric coupling is given by $H_{\epsilon}=\vec{d}\cdot \vec{\epsilon}=\sum_{i}e\vec{r}_{i}\cdot \vec{\epsilon}$. Here $e$ is the electron charge and $\vec{r}_{i}$ denotes the coordinate of the $i$th electron in the spin structure of single molecular magnet. In the following discussion, a typical case of triangular spin-$\frac 12$ molecular magnet $Cu_{3}$ \cite{Trif} is considered. At the low energies, the states of the system are represented by the quantum numbers of three spin-$\frac 12$ $\vec{s}_i$ labelling three $Cu^{2+}$ ions and the orbital states are quenched. The spin Hamiltonian of $Cu_{3}$ is described as,
\begin{equation}
\label{eq:2}
H_{0}=\sum_{i=1}^{3}J_{i,i+1}\vec{s}_{i}\cdot\vec{s}_{i+1}+\sum_{i=1}^{3}\vec{D}_{i,i+1}\cdot \vec{s}_{i}\times \vec{s}_{i+1}~,
\end{equation}
where the Heisenberg exchange couplings $J_{i,i+1}\thicksim 5~meV$ determine the gross structure of the energy spectrum and the Dzyaloshinskii-Moriya interactions $\vec{D}_{i,i+1}\thicksim 0.5~meV$ are one fine term. We can neglect the very small intrinsic deformation of the single molecular magnet. It is found that this spin Hamiltonian shows both spin-orbit interactions and the absence of spin inversion symmetry. To clearly describe the energy properties of $Cu_{3}$, we plot the configuration of the spin structure and energy levels described by Eq.~(2) in Figure 1. The ground state multiplet has the total spin $S=\frac 12$ and can be spanned by the opposite chirality states $\{ |\chi_{\pm},~S_z=\pm \frac 12\rangle\}$. $|\chi,~S_z=-\frac 12\rangle$ are the spin flipped states of $|\chi,~S_z=\frac 12\rangle$. Here $|\chi,~S_z\rangle$ are the simultaneous eigenvectors of chirality operator $C_{z}$ and total spin operator $S_{z}=\sum_{i}s_{i}^{z}$. The chirality operator is $C_{z}=\frac {4}{\sqrt{3}}\vec{s}_{1}\cdot (\vec{s}_{2}\times \vec{s}_{3})$. The operators $C_{\pm}$ reversing the chirality of the spin states satisfy that $C_{\pm}|\chi_{\mp} ,~S_z\rangle=|\chi_{\pm},~S_z\rangle$ and $C_{\pm}|\chi_{\pm},~S_z\rangle=0$. The chirality operators behave like the spin operators in the chiral space. In the ground state subspace spanned by $\{ |\chi_{\pm},~S_z=\frac 12\rangle\}$, the effective spin Hamiltonian $H_0$ can be expressed as
\begin{equation}
\label{eq:3}
H_0^{eff}=\frac 12 \omega_{so} C_{z}~,
\end{equation}
where $\omega_{so}$ is the effective spin-orbit interaction and can be calculated by the general symmetry group method \cite{Trif}. Applying a local and planar electric field $\vec{\epsilon}(t)$ with the frequency $\omega$, the perturbed spin-electric interaction Hamiltonian $H_{\epsilon}$ in the ground state subspace is described as,
\begin{widetext}
\begin{eqnarray}
\label{eq:4}
H_{\epsilon}^{eff}&=&\left(
\begin{array}{cc}
0~&\langle \chi_{+},\frac 12|H_{\epsilon}|\chi_{-},\frac 12\rangle\\
\langle \chi_{-},\frac 12|H_{\epsilon}|\chi_{+},\frac 12\rangle~& 0
\end{array}\right)=d\epsilon[e^{-i(\omega t+\beta)}C_{+}+e^{i(\omega t+\beta)}C_{-}]~.
\end{eqnarray}
\end{widetext}
By means of the planar coordinates $X=\sum_{i}x_i$ and $Y=\sum_{i}y_i$, the elements of the electric dipole matrix are calculated as $\langle \chi_{+},\frac 12|eX|\chi_{-},\frac 12\rangle=i\langle \chi_{+},\frac 12|eY|\chi_{-},\frac 12\rangle=d$ where $d\simeq 3.38\times 10^{-33}\textrm{C}~\textrm{m}$ ~\cite{Islam}~describes the strength of the electric dipole.
With the variation of the initial angle between the field $\vec{\epsilon}$ and the vector $\vec{r}_{12}$ from site $1$ to site $2$, we can reasonably adjust the phase angle $\beta=0$ \cite{Trif, Trif10}.

In the rotating frame with the electric field frequency $\omega$, the total Hamiltonian of the open system coupled to the environment can be written as
\begin{equation}
\label{eq:5}
H=H^{eff}+H_{E}+H_{I}~,
\end{equation}
where the effective spin Hamiltonian of the molecular magnet is transformed to
\begin{equation*}
H^{eff}=\frac 12 (\Delta_{so}C_{z}+d\cdot \epsilon C_{x})~.
\end{equation*}
The parameter $\Delta_{so}=\omega_{so}-\omega$ and the chirality operator $C_{x}=\frac 12(C_{+}+C_{-})$. In this case, we consider a finite-temperature environment modelled by a collection of bosons. The Hamiltonian of the thermal environment is expressed as
\begin{equation}
\label{eq:6}
H_{E}=\sum_{j}\omega_{j}b^{\dag}_{j}b_{j}~,
\end{equation}
where $b_{j}$ and $b^{\dag}_{j}$ are the annihilation and creation operator. The last term in Eq.~(5) denotes the weak interaction between the system and the environment, and can be written as
\begin{equation}
\label{eq:7}
H_{I}=\sum_{j}(g_{j}e^{-i\omega t}b^{\dag}_{j} C_{-}+g^{\ast}_{j}e^{i\omega t}b_{j} C_{+})~.
\end{equation}
The weak coupling $|g_{j}|=d\cdot \epsilon_{j}$ where $\epsilon_{j}$ represents the magnitude of the electromagnetic field of the $j$th mode with the frequency $\omega_{j}$. The thermal environment used above is fully characterized by the spectral density function $J(\omega^{\prime})=\sum_j | g_j|^{2}\delta(\omega^{\prime}-\omega_j)$. We need to mention that the present physical situation consists of a molecular magnet which is coupled to: (a) a cavity with a normal mode of frequency $\omega_0$ and a certain decay rate $\gamma$, and (b) an additional time dependent electric field of frequency $\omega$.

\subsection{\label{sec2:level22} Quantum master equation}

In the interaction representation, the decoherence of the low-energy spin state $\rho(t)$ can be approximately given by the second-order time-convolutionless master equation,
\begin{equation}
\label{eq:8}
\frac {d\rho(t)}{dt}=-\int_{0}^{t}dt_{1} \mathrm{Tr}_{E}[H^{\prime}_{I}(t),[H^{\prime}_{I}(t_1),\rho(t)\otimes\rho_{E}]],
\end{equation}
where $H^{\prime}_{I}(t)=e^{it(H^{eff}+H_{E})}H_{I}e^{-it(H^{eff}+H_{E})}$. The notation $\mathrm{Tr}_{E}$ is the partial trace over the freedom of the environment. It is assumed that the initial product state of the total system is $\rho_{tot}(0)=\rho(0)\otimes \rho_{E}$ where $\rho_{E}=\exp(-H_{E}/\kappa_{B}T)/\mathrm{Tr}[\exp(-H_{E}/\kappa_{B}T)]$ is the thermal equilibrium state of the environment and satisfies that $\mathrm{Tr}[H^{\prime}_{I}(t)\rho_{E}]=0$ \cite{Goan}. For the convenience, the Boltzmann constant $\kappa_{B}$ and Planck constant $h$ are also assumed to be one. The dimensionless low temperature condition of $T<\omega_{so}$ is considered here. In the following discussion, the temperature scale is set by the spin-orbit splitting.

To simplify the analytical calculation, the effective spin Hamiltonian can be diagonalized as $\bar{H}_{eff}=\frac {\omega_{s}}{2}\bar{C}_{z}$ where $\omega_{s}=\sqrt{\Delta^2_{so}+d^2\epsilon^2}$. The transformed chirality operator is $\bar{C}_{z}=U^{\dag}C_{z}U=|\Uparrow\rangle \langle \Uparrow|-|\Downarrow\rangle \langle \Downarrow|$ where the transformation operation is $U=|\psi_{+}\rangle \langle \Uparrow|+|\psi_{-}\rangle \langle \Downarrow|$. The eigenvector of $H^{eff}$ are $|\psi_{\pm}\rangle=\pm\sqrt{\delta_{\pm}}|\chi_{+},~\frac 12\rangle+\sqrt{\delta_{\mp}}|\chi_{-},~\frac 12\rangle$. The coefficients $\delta_{\pm}=(\omega_{s}\pm \Delta_{so})/2\omega_{s}$.  Then, the interaction Hamiltonian in this dressed state basis $|\Uparrow(\Downarrow)\rangle$ can be given by
\begin{equation}
\label{eq:9}
\bar{H}^{\prime}_{I}(t)=A^{\dag}(t)\otimes B(t)+A(t)\otimes B^{\dag}(t).
\end{equation}
Here $A(t)=\sum_{j}g_j e^{-i\omega_j t}b_j$ and $B^{\dag}(t)=\delta_0 \bar{C}_z+\delta_{+} e^{i\omega_s t}\bar{C}_{+}-\delta_{-} e^{-i\omega_s t}\bar{C}_{-}$ where $\delta_0=\sqrt{\delta_{+} \delta_{-}}$. The expression of the time-convolutionless master equation in the dressed state basis is obtained as
\begin{equation}
\label{eq:10}
\frac {d\bar{\rho}(t)}{dt}=-i[\bar{H}^{eff}+\bar{H}^{\prime},\bar{\rho}(t)]+\hat{L}[\bar{\rho}(t)]+\hat{O}[\bar{\rho}(t)].
\end{equation} Where $\bar{H}^{\prime}=\mathrm{Im}(\Gamma_0-\Gamma^{\prime}_0)\delta^2_{0}\bar{C}^2_{z}+\sum_{l=\pm}\mathrm{Im}(\Gamma_{l}-\Gamma^{\prime}_{l})\delta^2_{l}\bar{C}^{\dag}_{q}\bar{C}_{q}$. The parameters $\Gamma_{l}$ and $\Gamma^{\prime}_{l}(l=0,\pm)$ are determined by
\begin{eqnarray}
\label{eq:11}
\Gamma_{l}&=&\int_{0}^{t}dt_{1}\sum_{j}|g_j|^2\cdot \bar{n}_{j}e^{(\omega_j-\omega-q\omega_s)(t-t_1)} \nonumber \\
\Gamma^{\prime}_{l}&=&\int_{0}^{t}dt_{1}\sum_{j}|g_j|^2\cdot(\bar{n}_{j}+1) e^{(\omega_j-\omega-q\omega_s)(t-t_1)},
\end{eqnarray}
where $\bar{n}_j=(e^{\omega_j/T}-1)^{-1}$ is the mean number for the $j$th mode of the thermal environment at $T$ temperature.
The Lindblad superoperator in Eq. (10) is given by
\begin{equation}
\label{eq:12}
\hat{L}[\bar{\rho}(t)]=\sum_{m=z,\pm}\gamma_{m}(t)[\bar{C}_{m}\bar{\rho}\bar{C}^{\dag}_{m}-\frac 12\{\bar{C}^{\dag}_{m}\bar{C}_{m}, \bar{\rho} \}],
\end{equation}
where the decay rates are obtained as $\gamma_{z}(t)=2\delta^2_0\mathrm{Re}(\Gamma_0+\Gamma^{\prime}_0)$, $\gamma_{+}(t)=2\delta^{+}_0\mathrm{Re}(\Gamma_{+})+2\delta^{-}_0\mathrm{Re}(\Gamma^{\prime}_{-})$ and $\gamma_{-}(t)=2\delta^{-}_0\mathrm{Re}(\Gamma_{-})+2\delta^{+}_0\mathrm{Re}(\Gamma^{\prime}_{+})$. The notation $\mathrm{Im(Re)}$ denotes imaginary (real) part of a complex parameter.
The last term in Eq. (10) is very complicate,
\begin{widetext}
\begin{eqnarray}
\label{eq:13}
\hat{O}[\bar{\rho}(t)]&=&\Gamma_{0}\cdot[\delta_0\delta_{+}(\bar{C}_z\bar{\rho}\bar{C}_{-}-\bar{\rho}\bar{C}_{-}\bar{C}_z)-\delta_0\delta_{-}(\bar{C}_z\bar{\rho}\bar{C}_{+}-\bar{\rho}\bar{C}_{+}\bar{C}_z)]\nonumber \\
&+&\Gamma_{+}\cdot[\delta_0\delta_{+}(\bar{C}_{+}\bar{\rho}\bar{C}_{z}-\bar{\rho}\bar{C}_{z}\bar{C}_{+})-\delta_{+}\delta_{-}(\bar{C}_{+}\bar{\rho}\bar{C}_{+}-\bar{\rho}\bar{C}_{+}\bar{C}_{+})]\nonumber\\
&-&\Gamma_{-}\cdot[\delta_0\delta_{-}(\bar{C}_{-}\bar{\rho}\bar{C}_{z}-\bar{\rho}\bar{C}_{z}\bar{C}_{-})+\delta_{+}\delta_{-}(\bar{C}_{-}\bar{\rho}\bar{C}_{-}-\bar{\rho}\bar{C}_{-}\bar{C}_{-})]\nonumber\\
&+&\Gamma^{\prime}_{0}\cdot[\delta_0\delta_{+}(\bar{C}_{-}\bar{\rho}\bar{C}_{z}-\bar{\rho}\bar{C}_{z}\bar{C}_{-})-\delta_0\delta_{-}(\bar{C}_{+}\bar{\rho}\bar{C}_{z}-\bar{\rho}\bar{C}_{z}\bar{C}_{+})]\nonumber \\
&+&\Gamma^{\prime}_{+}\cdot[\delta_0\delta_{+}(\bar{C}_{z}\bar{\rho}\bar{C}_{+}-\bar{\rho}\bar{C}_{+}\bar{C}_{z})-\delta_{+}\delta_{-}(\bar{C}_{+}\bar{\rho}\bar{C}_{+}-\bar{\rho}\bar{C}_{+}\bar{C}_{+})]\nonumber\\
&-&\Gamma^{\prime}_{-}\cdot[\delta_0\delta_{-}(\bar{C}_{z}\bar{\rho}\bar{C}_{-}-\bar{\rho}\bar{C}_{-}\bar{C}_{z})+\delta_{+}\delta_{-}(\bar{C}_{-}\bar{\rho}\bar{C}_{-}-\bar{\rho}\bar{C}_{-}\bar{C}_{-})]+\mathrm{h.c.}
\end{eqnarray}
\end{widetext}
The notation $\mathrm{h.c.}$ represents Hermitian conjugate. The first part in Eq. (10) is the unitary one. $\bar{H}^{\prime}$ is the Lamb shift Hamiltonian and describes a small shift in the energy of the eigenvectors of $\bar{H}^{eff}$. The Lamb shift Hamiltonian has no qualitative effect on the decoherence of the system and may be neglected. Meanwhile, according to the calculation of the effective spin-electric coupling, the characteristic time for the low-energy molecular magnet of $Cu_{3}$ is about $\tau_{s}=\omega_{s}^{-1}\sim 10 \mathrm{ns}$ \cite{Islam} which is always much smaller than the correlation time of the thermal environment $\tau_{E}\sim 1 \mathrm{\mu s}$ \cite{Weiss99}. Under the condition of $\tau_{s}\ll\tau_{E}$, the influence of the last term in Eq. (9) on the decoherence of the chiral states is very small and usually negligible \cite{Breuer01}. This secular approximation is known as the rotating wave approximation which involves an averaging over the rapidly oscillating terms in the quantum master equation. This approximation is obtained by neglecting the high-frequency oscillating terms which are denoted by the term of Eq.~(13). The secular approximation is reasonable under the assumption of the weak coupling between the system and the environment. In the following, the parts of $\bar{H}^{eff}$ and Lindblad superoperator $\hat{L}[\bar{\rho}(t)]$ are dominant in the second-order time-convolutionless master equation describing the dynamics of spin states in an electrically driven single molecular magnet.

\section{Dynamics of electrically coupled spin states}

An efficient way to describe the dynamics of the low-energy spin chiral states is to analyze the evolution of the chirality spin polarization $P(t)=\mathrm{Tr}[\bar{\rho}(t)\bar{C}_z]$ where $\bar{\rho}(t)$ is the reduced density matrix of the single molecular magnet.
For the initial full polarization $P(0)=1$, the analytical expression of the chirality spin polarization at time $t$ are obtained as
\begin{equation}
\label{eq:14}
P(t)=\{1+\int_{0}^{t}dt_{1}\cdot e^{f(t_1)}[\gamma_{+}(t_1)-\gamma_{-}(t_1)] \}e^{-f(t)}~,
\end{equation}
where the function $f(t)=\int_{0}^{t}dt_{1}[\gamma_{+}(t_1)+\gamma_{-}(t_1)]$. From the above equation, we can easily find out the decaying rates of $\gamma_{\pm}$ determine the dynamics of chiral spin polarization. In the Markof approximation, the evolution of $P(t)$ presents the exponential decease because of $\vert \gamma_{\pm}\vert \rightarrow 0$. However, for a real physical model, it is interesting to characterize the behavior of the spin polarization. For an example, a traditional Lorentzian spectral density function is used to describe the thermal bosonic environment such as the quantized electromagnetic field inside a cavity. The weak interactions between the system and thermal environment are given by the spectral density function which is $J(\omega^{\prime})=\frac {\alpha^2\lambda^2}{2\pi[(\omega^{\prime}-\omega_{0})^2+\lambda^2]}$and the weak coupling constant $\alpha^2\ll \omega_{s}$. The correlation time scale is obtained as $\tau_{E}\thicksim\lambda^{-1}$ where $\lambda$ denotes the width of the distribution quantifying leakage of photons. $\omega_{0}$ is the center frequency of the electromagnetic field in the cavity. With the consideration of the quantum information processing, we focus on the dynamics of the spin states on the time scale of the correlation time of the environment.

Figure 2 demonstrates the effects of the local electric field on the decoherence of the chiral states at the time interval of $t\sim \tau_{E}$. For the small values of $\frac {\Delta_{so}}{\omega_{s}}=\frac {\omega_{so}-\omega}{\sqrt{d^2\epsilon^2+(\omega_{so}-\omega)^2}}$, the decay of the chiral spin polarization can be restrained to some extent. This point shows that the increase of the strength of the local electric field is useful for the control of the coherence. It is found out that the coherence can also be improved by the manipulation of the frequency $\omega$ of the electric field which is almost resonant with the transition frequency $\omega_{so}$ of the spin-orbit interaction. The suppressed and revived behavior of the polarization manifest the information exchange between the system and thermal environment. This rapid non-Markovian oscillation is mainly induced by the weak coupling to the memory environment.

To physically explain the dynamics of the chiral states, we also calculate the non-Markovian behavior of the decay rates for the quantum master equation on the time scale of $\tau_{E}$ in Figs. 3. The time evolution of the decay rates $\gamma_{m}(t),(m=\pm)$ is plotted for different temperatures. The $\gamma_{m}(t),(m=\pm)$ is plotted in Figs. 3(a) and 3(b) for the small value of $\frac {\omega_0-\omega}{\lambda}=0.1$. The values of $\gamma_{\pm}$ always oscillate between some positive and negative values with slightly damping rate. The amplitude of $\gamma_{\pm}(t)$ for $T=0$ is smaller than that for $T=1$. In non-Markovian quantum jumps formalism, negative values are regarded as the occurrence of reversed quantum jumps which can indicate the non-Markovian dynamics induced by the environmental memory. The memory effects describe the exchange of energy and information between the system and environment. The $\gamma_{m}(t),(m=\pm)$ is plotted in Figs. 3(c) and 3(d) for large value of $\frac {\omega_0-\omega}{\lambda}=10$. The curve of $T=0$ is almost the same as that of $T=1$. That is, the influence of low temperatures on the dynamics of the decay rates is almost negligible in this case. According to ~\cite{Breuer01}, the larger values of $\frac {\omega_0-\omega}{\lambda}$ represent the smaller effective coupling between the system and environment. In this very weak coupling case, the effects of temperatures on the decoherence are also very small. For the long time limit, the values of the decay rates approach to some steady value which is infinitely close to zero. This means that the non-Markovian dynamics on the time scale of $\tau_E$ is reduced to the Markovian one on the long time scale. Therefore, the dynamics of the spin states on the time scale of the correlation time of the environment can have the prominent effect on the rapid control of the single molecular magnet.

To further elaborate the validity of the weak coupling approximation mentioned above, we can equivalently map the model into another one which is widely used in the solid-state systems~\cite{Thorwart04, Thorwart042, Zueco08}. This new dissipation model includes an effective two-level system like the single molecular magnet which is weakly coupled to the normal cavity mode with the frequency $\omega_0$ and operator $a$. The weak interaction strength between the two-level system and the normal mode is denoted by $g$. The new bath is restricted to the remaining oscillator modes coupled to the normal mode. The corresponding Hamiltonian is written as,
\begin{equation}
H=H_S+(a^{\dag}+a)\sum_j \nu_j(\tilde{b}_j^{\dag}+\tilde{b}_j)+\sum_j\tilde{\omega}_j\tilde{b}_j^{\dag}\tilde{b}_j,
\end{equation}
where $H_S=H^{eff}+g(a C_{+}+a^{\dag}C_{-})+\omega_0a^{\dag}a+(a^{\dag}+a)^2\sum_j\nu_j^2/2\tilde{\omega}_j$. The bath is described by the Ohmic spectral density function $J_{Ohm}(\omega^{\prime})=\sum_j|\nu_j|^2\delta(\omega^{\prime}-\tilde{\omega}_j)=\gamma\omega^{\prime}\exp(-\omega^{\prime}/\omega_c)$ where the decay rate $\gamma$ is small and the cutoff frequency is $\omega_c$. In this case, we also consider the spin-boson model given by Eq.~(5) with the effective spectral density function of $J_{eff}(\omega^{\prime})=\frac {2\alpha\omega^{\prime}\omega_0^4}{(\omega_0^2-\omega^{\prime 2})^2+(2\pi\gamma\omega^{\prime}\omega_0)^2}$. The relation between $g$ and weak coupling strength $\alpha$ follows as $\alpha=8\gamma\frac {g^2}{\omega_0}$  According to the results of~\cite{Thorwart04, Thorwart042}, the weak coupling approximation is applicable when the coupling strength $g\ll \gamma$. The dynamics of the spin states can also be evaluated by the method introduced by~\cite{Zueco08}. Figure 4 shows the dynamics of spin polarization when the new dissipation model is used. There exist the small revivals of the polarization which are largely suppressed in comparison with the results of Figure 2.

In respect to quantum information processing, the stability of the information storage needs to be analyzed when encoding qubits in single molecular magnet are electric-controllable in the thermal environment. From the perspective of von Neumann entropy, the pointer state \cite{Paz, Khodjastech} can be defined as one initial state which becomes minimally entangled with the environment during the evolution. The study of the pointer state can help us to understand the effects of the decoherence on quantum information processing. The entropy for the reduced density matrix $\bar{\rho}(t)$ of the non-Markovian decoherence can be written as,
\begin{equation}
\label{eq:15}
E(t)=-\mathrm{Tr}[\bar{\rho}(t)\ln\bar{\rho}(t)]=\sum_{i=1,2}u_{i}\ln u_{i}~,
\end{equation}
where $u_{i}$ is the $j$-th eigenvalues of $\bar{\rho}$. We can use the Bloch vector to describe the expression of the chiral spin states, $\bar{\rho}=\frac {I+\vec{V}\cdot \vec{C}}{2}$ where the Bloch vector $\vec{V}=(\sin \theta \cos \phi, \sin \theta
\sin \phi, \cos \theta)$ and $\theta\in [0,\pi],\phi\in [0,2\pi]$.

At $T=0$, the entropy $E$ during the decoherence is plotted in Figs. 5. It is seen that the values of the entropy are always increased with the time in Fig. 5(a). The larger values of the entropy denote the more entanglement between the system and environment. When the initial state is at $\theta=\pi$, the values of $E$ almost remain the minimal ones in the evolution. Therefore, the pointer state for the thermal environment with $T=0$ is the state of $\theta=\theta_{p}=\pi$ which is almost the ground state $|\psi_{-}\rangle$. In fact, the pointer state is determined by the properties of the environment \cite{Paz, Khodjastech}. At low temperature of $T=1$, the dynamics of the entropy is shown in Fig. 5(b). It is clearly seen that the values of the entropy for $T=1$ are always increased more quickly than those of $T=0$. This means that the single molecular magnet is easily entangled with the environment at $T\neq 0$. At a finite temperature $T=1$, the pointer state with the initial angle $\theta_{p}$ are varied with the parameter $\frac {\omega_0-\omega}{\lambda}$ in Figure 6. For large value of $\frac {\omega_0-\omega}{\lambda}\geq 10$, the pointer state is approximately the ground state because of the very small effective coupling between the system and the environment.

\section{DISCUSSION}

The decoherence of the low-energy spin states in an electrically driven single molecular magnet weakly coupled to a thermal environment is investigated. By means of the time-convolutionless non-Markovian master equation, the reduced density matrix for the spin states can be derived in the condition of $\tau_{s}\ll \tau_{E}$. In regard to the Lorentzian environment, the oscillations of the decay rates between positive values and negative ones appear. This phenomenon indicates the memory effects from the non-Markovian environment. The rapid non-Markovian decoherence of the Bloch vector occurs due to the quick exchange of energy and information between the system and the environment. The decoherence can be efficiently suppressed by adjusting the electric field and the parameters of the environmental spectral density function. In quantum information processing, the selection of the pointer states can be determined by the properties of the environment.

\section{ACKNOWLEDGEMENT}

This work is supported by the National Natural Science Foundation of China under Grant No. 10904104, No. 11074184 and No. 11174363.

\newpage

{\Large \bf Figure Captions}

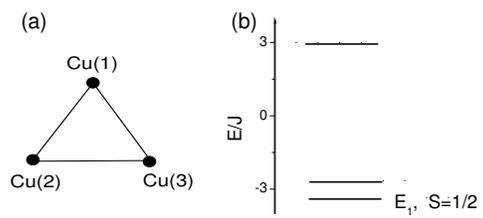
{\bf Fig. 1}

(a) The configuration of spin structure is shown; (b) The energy level scheme is plotted for the spin-orbit interaction $D/J=0.1$ and $\vec{D}=(0,0,D)$.

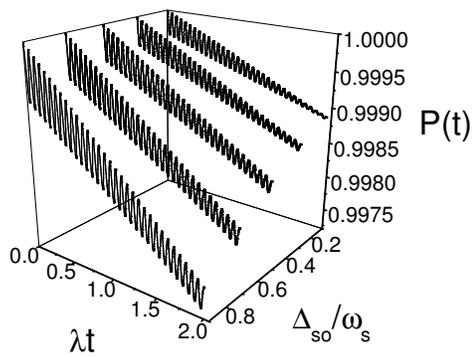
{\bf Fig. 2}

The dependence of the decoherence of the spin states on the electric field is plotted for $\frac {\omega_{s}}{\lambda}=100$, $\frac {\omega_0-\omega}{\lambda}=0.1$ and $T=1$. The parameter $\frac {\Delta_{so}}{\omega_{s}}$ can be modified with the frequency $\omega$ or the strength $\epsilon$ of the electric field.

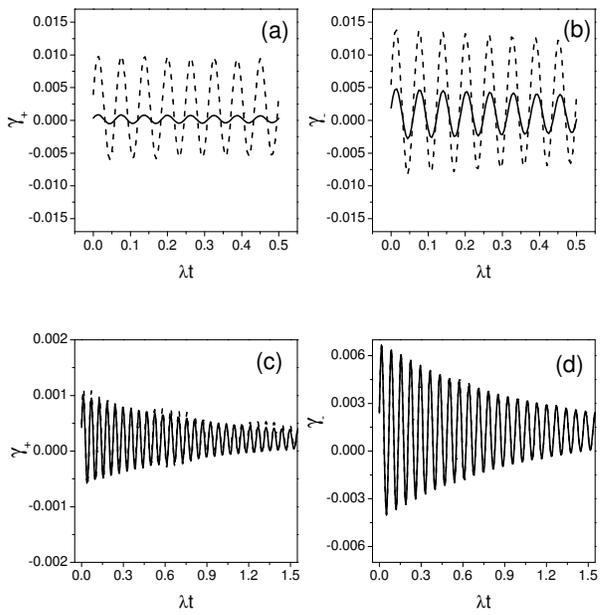
{\bf Fig. 3}

The time evolution of the decay rates $\gamma_{m}(t),(m=\pm)$ is plotted as a function of the scaled time $\lambda t$ when the parameters are $\frac {\omega_{s}}{\lambda}=100$ and $\frac {\Delta_{so}}{\omega_{s}}=0.4$. For (a) and (b), the parameter is $\frac {\omega_0-\omega}{\lambda}=0.1$ while for (c) and (d), the parameter is $\frac {\omega_0-\omega}{\lambda}=10$. The solid lines denote the case of $T=0$ and dashed ones represent that of $T=1$.

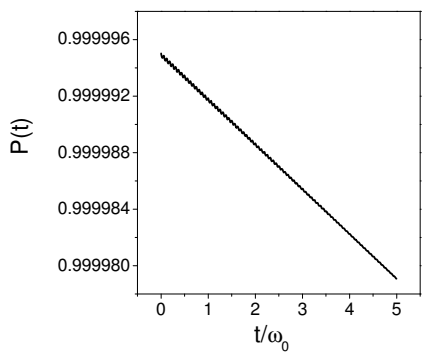
{\bf Fig. 4}

The dynamics of the spin polarization at temperature $T=1$ is plotted using the new dissipation model for $\gamma=0.1$, $g=0.01\omega_0$, $\omega_s=100\omega_0$, $\omega=0.9\omega_0$ and and $\frac {\Delta_{so}}{\omega_{s}}=0.4$.

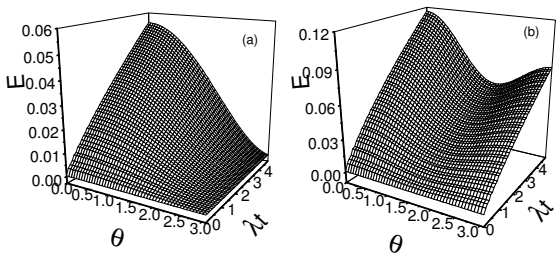
{\bf Fig. 5}

(a). The dynamics of the von Neumann entropy $E$ is plotted at $T=0$. The initial state are changed with $\theta$ and the parameters are $\frac {\omega_{s}}{\lambda}=100$, $\frac {\omega_0-\omega}{\lambda}=0.1$ and $\frac {\Delta_{so}}{\omega_{s}}=0.9$. (b). The dynamics of the von Neumann entropy at $T=1$ is plotted for $\frac {\omega_{s}}{\lambda}=100$ and $\frac {\Delta_{so}}{\omega_{s}}=0.9$. The initial state are changed with $\theta$ when $\frac {\omega_0-\omega}{\lambda}=10$.

{\bf Figs. 6}

The pointer state represented by $\theta_{p}$ is plotted as a function of the environment parameter in the condition of $T=1$, $\frac {\omega_{s}}{\lambda}=100$ and $\frac {\Delta_{so}}{\omega_{s}}=0.9$.

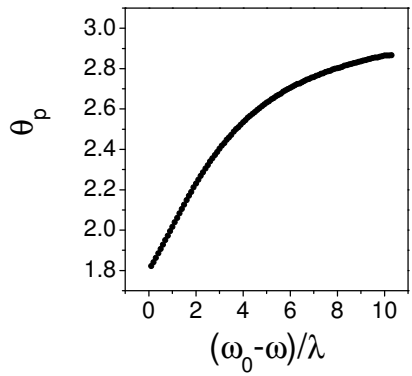
\end{document}